\documentclass[5p, times, sort&compress]{elsarticle}

\usepackage{ecrc}

\volume{00}

\firstpage{1}

\journalname{Photonics and Nanostructures - Fundamentals and Applications}

\runauth{D.A. Kuzmin at al.}

\jid{pnfa}

\jnltitlelogo{Photonics and Nanostructures - Fundamentals and Applications}

\CopyrightLine{2014}{Published by Elsevier Ltd.}

\usepackage{amssymb}
\usepackage{amsmath}

\usepackage[figuresright]{rotating}

\begin{document}

\begin{frontmatter}



\dochead{}

\title{Electromagnetic waves reflection, transmission and absorption  by graphene - magnetic semiconductor - graphene sandwich-structure in magnetic field: Faraday geometry}


\author[1]{Dmitry A. Kuzmin\corref{cor1}}
\ead{kuzminda89@gmail.com}
\author[1]{Igor V. Bychkov}
\ead{bychkov@csu.ru}
\author[2]{Vladimir G. Shavrov}

\address[1]{Chelyabinsk State University, 454001, Chelyabinsk, Br. Kashirinyh Street, 129, Russia}
\address[2]{Kotelnikov Institute of Radioengineering and Electronics of the Russian Academy of Sciences, 125009, Moscow, Mokhovaya Street, 11-7, Russia}

\cortext[cor1]{Corresponding author}

\begin{abstract}
Electrodynamic properties of the graphene - magnetic semiconductor - graphene sandwich-structure have been investigated theoretically with taking into account the dissipation processes. Influence of graphene layers on electromagnetic waves propagation in graphene - semi-infinte magnetic semiconductor and graphene - magnetic semiconductor - graphene sandwich-structure has been analyzed.  Frequency and field dependences of the reflectance, transmittance and absorbtance of electromagnetic waves by such structure have been calculated. The size effects associated with the thickness of the structure have been analyzed. The possibility of efficient control of electrodynamic properties of graphene - magnetic semiconductor - graphene sandwich structure by an external magnetic field has been shown.
\end{abstract}


\end{frontmatter}


\section{Introduction}
\label{intro}
Nowadays graphene (two dimensional lattice of carbon atoms) attracts researchers' attention with their special properties \cite{1, 2, 3, 4}, including electrodynamical ones. So, for example, exciting of the surface plasmons in the graphene layers has been investigated theoretically \cite{5, 6, 7, 8, 9, 10, 11} and experimentally \cite{12, 13}. These studies showed that both TE- and TM- polarized plasmons can exist in graphene. What is more, it is possible to control their dispersion characteristics by applying a voltage. Recently, the possibility of control of the hybrid surface waves in graphene placed between two dielectrics by applying an external magnetic field \cite{14} and the waveguide properties of the sandwich-structure graphene-dielectric graphene \cite{15} have been studied; the existence of TE- and TM- polarized plasmon modes of THz frequencies, which are mixed in the presence of a magnetic field, has been shown. The possibility of existence of waveguide modes with the negative group speed has been demonstrated as well. In work \cite{16} the potential of creating of the hyperbolic metamaterial based on graphene – dielectric multilayer structure has been shown. The authors also noted that such metamaterial cannot be attributed to an elliptical or hyperbolic type in the presence of magnetic field due to the Hall effect in graphene, that allows one to control its properties. Reflection and transmission of electromagnetic waves by the graphene layer and graphene superlattice have been investigated in details in \cite{17, 18}.

Despite the large number of studies, the authors are usually limited by investigation of a non-magnetic dielectric medium, where graphene is placed. It is very interesting to study the dynamic characteristics of graphene-based structures with more complex materials. Recently, metamaterial
composed of periodic stacking of graphene - liquid crystal layers has been proposed for far-infrared frequencies \cite{19}. The permeability of the liquid crystal and the surface conductivity of the graphene sheets are the tunable parameters. So, the optical properties of the structure can be controlled and the metamaterial is able to show both the elliptic and hyperbolic dispersions. A magnetic semiconductor could be another example of  material with tunable permeability and permittivity. Semiconductor superlattices studied for a long time (see, for example, \cite{20} and Refs. therein); however, they are still of 
the interest \cite{21, 22}. Frequency dispersion of the permittivity is one of the significant differences of semiconductor from dielectrics; the plasma waves can be excited in the semiconductor structures. When the semiconductor is placed in an external magnetic field, the helicons can to propagate in the material. Their properties depend on the magnetic field value. In its turn, the magnetic semiconductors have a number of specific features. For example, they may have a large magnetoresistance \cite{23, 24}, magnetooptical properties \cite{25, 26}, etc. Thus, the electrodynamic properties of graphene-magnetic semiconductor-based structures can be quite interesting. This paper is devoted to investigation of the reflection, transmission and absorption of electromagnetic waves by graphene - magnetic semiconductor - graphene sandwich-structure placed in an external magnetic field. All over the paper CGS units are used.


\section{Theory}
\label{theory}
\begin{figure} [h!]
\center{\includegraphics[width=90mm]{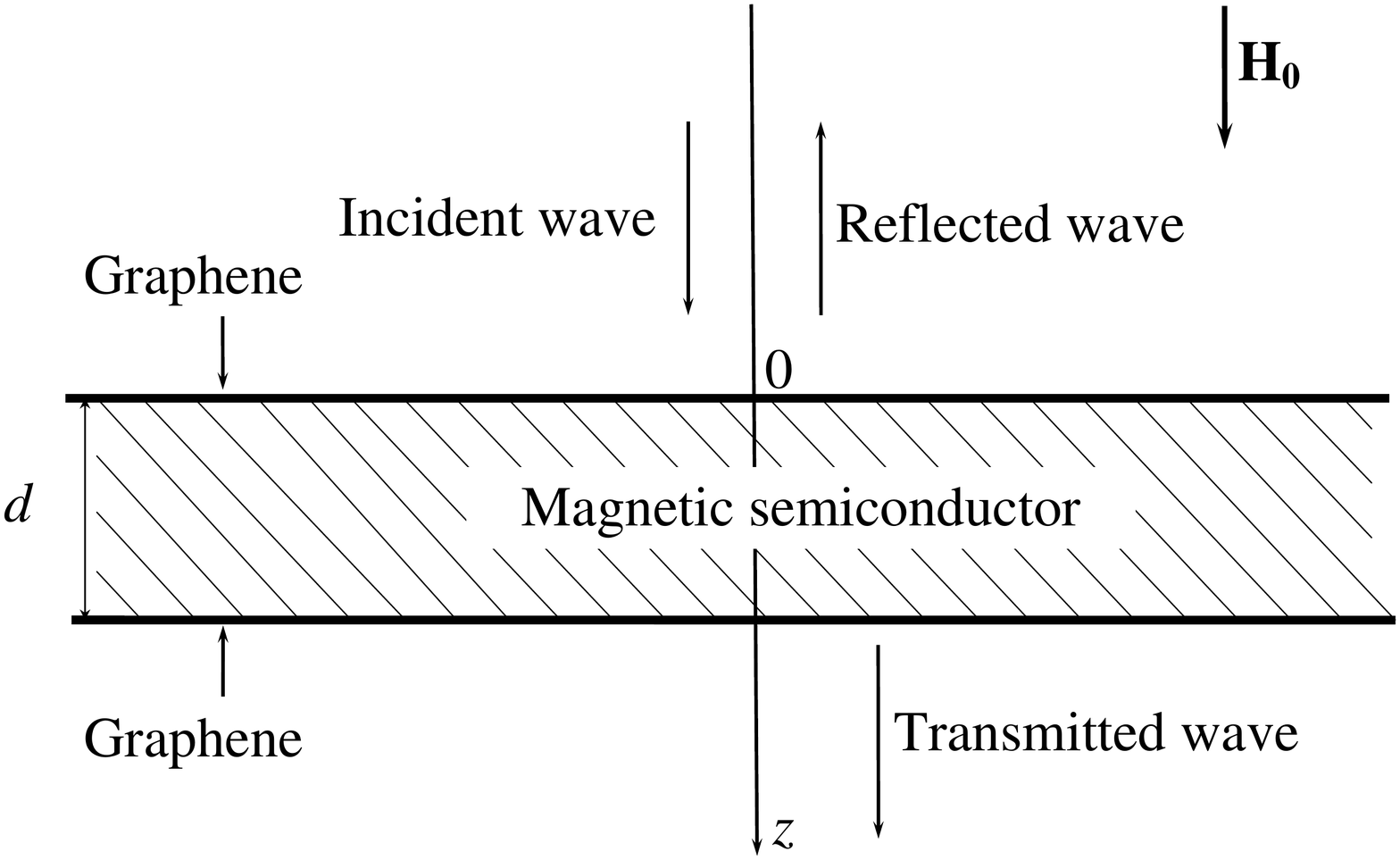}}
\caption{Geometry of the problem.}
\label{ris:geometry}
\end{figure}

Let us consider the graphene - magnetic semiconductor - graphene sandwich-structure placed in an external magnetic field $\mathbf{H_0}$, which is directed perpendicular to the structures' surface, in vacuum. Suppose that linearly polarized plane harmonic electromagnetic wave with time dependence of exp$(-i \omega t)$ ($\omega = 2\pi f$ is an angular frequency, $f$ is a linear frequency in Hz; we will use $f$ in figures) is normally incident in the surface of the structure, shown in Figure \ref{ris:geometry} (Faraday geometry). It is sufficient to consider an electromagnetic wave polarized along only one axis due to the axial symmetry of the problem. The coordinate axes are chosen so that the $z$-axis coincides with the direction of the external magnetic field, e.g. $\mathbf{H_0} = (0, 0, H_0)$. The thickness of the magnetic semiconductor is denoted $d$.

For solving this problem, one has to know the characteristics of each component of the structure. For the magnetic semiconductor such characteristics are the tensors of the permeability $\hat{\mu }$ and the permittivity $\hat{\varepsilon }$. The permeability tensor of the magnetic semiconductor placed in an external magnetic field can be described as following:
\begin{equation}\label{eq:mu_tens}
\begin{matrix}
  \hat{\mu }=\left( \begin{matrix}
   {{\mu }_{\bot }} & i{{\mu }_{a}} & 0  \\
   -i{{\mu }_{a}} & {{\mu }_{\bot }} & 0  \\
   0 & 0 & {{\mu }_{\parallel }}  \\
\end{matrix} \right); \\ 
  {{\mu }_{\bot }}=1+\frac{{{\omega }_{M}}\left( {{\omega }_{H}}-i\alpha \omega  \right)}{\omega _{H}^{2}-\left( 1+{{\alpha }^{2}} \right){{\omega }^{2}}-2i\alpha \omega {{\omega }_{H}}};\ {{\mu }_{a}}=\frac{-{{\omega }_{M}}\omega }{\omega _{H}^{2}-\left( 1+{{\alpha }^{2}} \right){{\omega }^{2}}-2i\alpha \omega {{\omega }_{H}}}; \\ 
  {{\mu }_{\parallel }}=1-\frac{i\alpha {{\omega }_{M}}}{\omega +i\alpha {{\omega }_{H}}}. \\ 
\end{matrix}
\end{equation}
	
In (\ref{eq:mu_tens}) we used the following notation: ${\omega }_{H}=g{{H}_{0}}$, ${\omega }_{M}=4\pi g{{M}_{0}}$, $g$ is the gyromagnetic ratio, $M_0$ is the saturation magnetization, $\alpha $ is the damping parameter.
	
The permittivity tensor has the usual form \cite{20}:
\begin{equation}\label{eq:eps_tens}
\begin{matrix}
  \hat{\varepsilon }=\left( \begin{matrix}
   {{\varepsilon }_{\bot }} & i{{\varepsilon }_{a}} & 0  \\
   -i{{\varepsilon }_{a}} & {{\varepsilon }_{\bot }} & 0  \\
   0 & 0 & {{\varepsilon }_{\parallel }}  \\
\end{matrix} \right); \\ 
  {{\varepsilon }_{\bot }}={{\varepsilon }_{0}}\left( 1-\frac{\omega _{p}^{2}\left( \omega +i\nu  \right)}{\omega \left[ {{\left( \omega +i\nu  \right)}^{2}}-\omega _{c}^{2} \right]} \right);\ {{\varepsilon }_{a}}={{\varepsilon }_{0}}\frac{\omega _{p}^{2}{{\omega }_{c}}}{\omega \left[ {{\left( \omega +i\nu  \right)}^{2}}-\omega _{c}^{2} \right]}; \\ 
  {{\varepsilon }_{\parallel }}={{\varepsilon }_{0}}\left( 1-\frac{\omega _{p}^{2}}{\omega \left( \omega +i\nu  \right)} \right). \\ 
\end{matrix}
\end{equation}

Here, ${{\varepsilon }_{0}}$ is the lattice caused part of the permittivity, ${{\omega }_{p}}=\sqrt{{4\pi {{n}_{s}}{{e}^{2}}}/{{{m}^{*}}}\;}$ and ${{\omega }_{c}}={e{{H}_{0}}}/{{{m}^{*}}c}\;$ are the plasma and the cyclotron frequencies, consequently, $e$ and ${{m}^{*}}$ are the charge and the effective mass of the charge carriers, $c$ is the speed of light in vacuum, ${{n}_{s}}$ is the carriers density, $\nu$ is the effective collision rate. 

Graphene can be represented as a conductive surface \cite{5} with the frequency dependent tensor of conductivity $\hat{\sigma }$, which components have been obtained in \cite{27, 28}
\begin{equation}\label{eq:sigma_tens}
\begin{matrix}
  \hat{\sigma }=\left( \begin{matrix}
   {{\sigma }_{0}} & {{\sigma }_{H}}  \\
   -{{\sigma }_{H}} & {{\sigma }_{0}}  \\
\end{matrix} \right); \\ 
  {{\sigma }_{0}}=\frac{{{e}^{2}}v_{F}^{2}\left| e{{H}_{0}} \right|\left( \hbar \omega +2i\Gamma  \right)}{i\pi c}\times  \\ 
  \times \sum\limits_{n}{\left\{ \begin{matrix}
  & \frac{\left[ {{n}_{F}}\left( {{M}_{n}} \right)-{{n}_{F}}\left( {{M}_{n+1}} \right) \right]-\left[ {{n}_{F}}\left( -{{M}_{n}} \right)-{{n}_{F}}\left( -{{M}_{n+1}} \right) \right]}{{{\left( {{M}_{n+1}}-{{M}_{n}} \right)}^{3}}-{{\left( \hbar \omega +2i\Gamma  \right)}^{2}}\left( {{M}_{n+1}}-{{M}_{n}} \right)}+ \\ 
 & +\frac{\left[ {{n}_{F}}\left( -{{M}_{n}} \right)-{{n}_{F}}\left( {{M}_{n+1}} \right) \right]-\left[ {{n}_{F}}\left( {{M}_{n}} \right)-{{n}_{F}}\left( -{{M}_{n+1}} \right) \right]}{{{\left( {{M}_{n+1}}+{{M}_{n}} \right)}^{3}}-{{\left( \hbar \omega +2i\Gamma  \right)}^{2}}\left( {{M}_{n+1}}+{{M}_{n}} \right)} \\ 
\end{matrix} \right\}}; \\ 
  {{\sigma }_{H}}=-\frac{{{e}^{2}}v_{F}^{2}e{{H}_{0}}}{\pi c}\times  \\ 
  \times \sum\limits_{n}{\left\{ \begin{matrix}
  & \left\{ \left[ {{n}_{F}}\left( {{M}_{n}} \right)-{{n}_{F}}\left( {{M}_{n+1}} \right) \right]+\left[ {{n}_{F}}\left( -{{M}_{n}} \right)-{{n}_{F}}\left( -{{M}_{n+1}} \right) \right] \right\}\times  \\ 
 & \times \frac{2\left[ M_{n}^{2}+M_{n+1}^{2}-{{\left( \hbar \omega +2i\Gamma  \right)}^{2}} \right]}{{{\left[ M_{n}^{2}+M_{n+1}^{2}-{{\left( \hbar \omega +2i\Gamma  \right)}^{2}} \right]}^{2}}-4M_{n}^{2}M_{n+1}^{2}} \\ 
\end{matrix} \right\}} \\ 
\end{matrix}
\end{equation}

Here, ${{M}_{n}}={{v}_{F}}{{\left( {2ne{{H}_{0}}}/{c}\; \right)}^{{1}/{2}\;}}$ is energy of the corresponding Landau level, ${{n}_{F}}$ is the function of Fermi-Dirac distribution, ${{v}_{F}}$ is the velocity of electrons on the Fermi surface, ${\Gamma }/{\hbar }\;$ is the scattering rate.
	
For solving the problem one has to use the system of Maxwell's equations
\begin{equation}\label{eq:maxwell}
rot\mathbf{E}=-{{c}^{-1}}{\partial \mathbf{B}}/{\partial t;\quad rot\mathbf{H}}\;={{c}^{-1}}{\partial \mathbf{D}}/{\partial t}\;
\end{equation}
with the material equations
\begin{equation}\label{eq:material}
\mathbf{D}=\hat{\varepsilon }\mathbf{E};\quad \mathbf{B}=\hat{\mu }\mathbf{H};\quad \mathbf{j}=\hat{\sigma }\mathbf{E}
\end{equation}
and the boundary conditions
\begin{equation}\label{eq:boundary}
\left( {{\mathbf{E}}_{2}}-{{\mathbf{E}}_{1}} \right)\times {{\mathbf{n}}_{12}}=0;\quad \left( {{\mathbf{H}}_{2}}-{{\mathbf{H}}_{1}} \right)\times {{\mathbf{n}}_{12}}={4\pi \mathbf{j}}/{c}\;
\end{equation}
where, indexes "1" and "2" mean the fields in the first and the second medium, ${{\mathbf{n}}_{12}}$ is the normal vector to the partition surface directed from the first medium to the second one, $\mathbf{j}$ is the density of the surface current of the graphene layer, $\mathbf{E}, \mathbf{D}$ ($\mathbf{H}, \mathbf{B}$) are electric (magnetic) field strength and induction, consequently.
	
Dispersion equation of the magnetic semiconductor has a usual for bigyrotropic medium form \cite{29}:
\begin{equation}\label{eq:dispertion}
{{k}_{\pm }}={{k}_{0}}\sqrt{\left( {{\varepsilon }_{\bot }}\pm {{\varepsilon }_{a}} \right)\left( {{\mu }_{\bot }}\pm {{\mu }_{a}} \right)}
\end{equation}
where ${{k}_{0}}={\omega }/{c}\;$ is the wave number of electromagnetic wave in vacuum, indexes "$+$" and "$-$" correspond to the right- and left- polarized waves.

Solving the system of equations \eqref{eq:maxwell}-\eqref{eq:boundary} with the dispersion equation \eqref{eq:dispertion} we obtain the amplitudes of reflected and transmitted waves. Then reflectance $R$ and transmittance $T$ can be found:
\begin{equation}\label{eq:R_T}
R=\frac{{{\left| {{E}_{xR}} \right|}^{2}}+{{\left| {{E}_{yR}} \right|}^{2}}}{{{\left| {{E}_{x0}} \right|}^{2}}+{{\left| {{E}_{y0}} \right|}^{2}}};\quad T=\frac{{{\left| {{E}_{xT}} \right|}^{2}}+{{\left| {{E}_{yT}} \right|}^{2}}}{{{\left| {{E}_{x0}} \right|}^{2}}+{{\left| {{E}_{y0}} \right|}^{2}}},
\end{equation}
where indexes "$R$" and "$T$" denote amplitudes of reflected and transmitted waves, consequently, index "$0$" denotes amplitude of incident wave. Obtaining their values we define the absorptance $A$
\begin{equation}\label{eq:A}
A=1-R-T
\end{equation}

Expressions for reflectance, transmittance and absorptance are very cumbersome; therefore limit ourselves to the numerical analysis.

\section{Material parameters for numerical simulation}
\label{parameters}

The ferrites, usually used for microwave engineering as well as in many experiments on ferromagnetic resonance and spin-waves are usually semiconductors with such small conductivity, that the effect of charge carriers may be neglected \cite{29}. But there are some ferrites with larger conductivity (e.g., \cite{30}) and such an effect become significant. Moreover, there are magnetic semiconductors \cite{31}, which combine magnetic ordering with conductivity as large as for "good" semiconductors. The other way to combine both semiconductor and ferromagnetism is to dope "classical" semiconductor by any magnetic element (for example, by Mn). This way has led to the fact that over the past fifth teen years, the field of ferromagnetism in diluted magnetic semiconductors and oxides has developed into an important branch of material science \cite{32}. All these material have a very wide range of parameters.

For example, CdCr$_2$Se$_4$ is well studied magnetic semiconductor. CdCr$_2$Se$_4$ can be n- and p-type magnetic semiconductor (depending on the method of preparation \cite{33}), it has a Curie temperature $T_c = 130$ K  \cite{34}, spontaneous magnetization $M_s = 225 - 245$ kA/m at temperature $T = 77$ K  \cite{35}, mobility about $100$ cm$^2$/V$\cdot$s  \cite{36}, resistivity $~ 10^4$ Ohm$\cdot$cm  \cite{37}, carrier concentration $n \sim 10^{16} - 10^{17}$ cm$^{-3}$ and narrow FMR line $2 \Delta H = 1.6$ Oe  \cite{36}. From this data we can estimate some parameters, needed for our calculation: $ \alpha = \Delta H/H \sim 0.001$ in magnetic field of kOe values; $\omega_p \sim 10^{12} - 10^{14}$ s$^{-1}$,  $ \nu \sim 10^{13}$ s$^{-1}$. Properties of diluted magnetic semiconductor Ti$_{1-x}$Co$_x$O$_{2-δ}$ may change significantly depending on the method of preparation. It may have a high Curie temperature $ \sim 600$ K \cite{38}, the charge carriers concentration usually is from $ \sim 10^{18} $ cm $^{-3}$ to $ \sim 10^{22}$ cm$^{-3}$ \cite{39}, electric resistance can vary from $10^{-4}$ to $10^6$ Ohm$\cdot$cm \cite{40, 41}, saturation magnetization could be about 10 kA/m \cite{42}, data on FMR research of this material, unfortunately, is absent. For this data $\omega_p \sim 10^{14} - 10^{16}$ s$^{-1}$,  $ \nu \sim 10^{11} - 10^{18}$ s$^{-1}$. Si$_{1-x}$Mn$_x$ is very interesting diluted magnetic semiconductor for our investigation due to possibility of epitaxial growth of grapheme on silicone substrate \cite{43}. This material may also be room temperature magnetic p-type semiconductor with hole concentration $ \sim 10^{22} $ cm$^{-3}$, resistivity $ \sim 10^{-4}$ Ohm$\cdot$cm, and saturation magnetization 10 - 400 G (depending on the concentration of Mn, temperature and method of preparation) \cite{44, 45}. This data lets us to estimate $\omega_p \sim 10^{16}$ s$^{-1}$, $ \nu \sim 10^{15}$ s$^{-1}$. There are many other diluted magnetic semiconductors having high $T_c$ \cite{46, 47, 48}.

 For the numerical simulation we will use the parameters of some magnetic semiconductor, which fall within the range mentioned above:
\begin{equation}\label{eq:parameters}
\begin{matrix}
  & {{M}_{0}}=160\ \text{G}\text{, }\alpha \text{ = 0}\text{.05}\text{, g =1}\text{.75}\cdot \text{1}{{\text{0}}^{7}}\text{ O}{{\text{e}}^{-1}}{{\text{s}}^{-1}}, \\ 
 & {{\varepsilon }_{0}}=17.8,\text{ }{{m}^{*}}=0.1{{m}_{e}},\text{ }{{\omega }_{p}}={{10}^{12}}{{\text{s}}^{-1}},\text{ }\nu \text{ = 1}{{\text{0}}^{11}}{{\text{s}}^{-1}}. \\ 
\end{matrix}
\end{equation}
	
For simulation of the graphene properties will use parameters ${{v}_{F}}={{10}^{8}}$ cm/s, $\Gamma = 2 \cdot 10^{-15}$ erg, value of the chemical potential ${{\mu }_{chem}}$ from the Fermi-Dirac distribution function depend on temperature $T$ and carriers' density in graphene ${{n}_{0}}$. How it has been shown in \cite{49}, for near-room temperatures ($T$ =300 K) and carriers' density ${{n}_{0}}\sim 10^{11}$ cm$^{-2}$ value of chemical potential is ${{\mu }_{chem}}\sim  3.5 \cdot 10^{-14}$ erg.

\section{Influence of graphene layers on electrodynamics of the structure}
\label{influence}

The graphene thickness is of the order of atomic distances. One thus expects that graphene will have no effect in refletance, transmittance and absorptance of electromagnetic waves in frequency range at least up to thausends of THz. Let us show now that presence of graphene layers lead to some effects even at GHz frequencies. Graphene can be represented as a conductive surface \cite{5} with the frequency dependent tensor of conductivity $\hat{\sigma }$. Due to small thickness of graphene layer we will take into account its presence only like a boundary condition (surface current $\mathbf{j}$ in \eqref{eq:boundary}).

\begin{figure} [h!]
\center{\includegraphics[width=90mm]{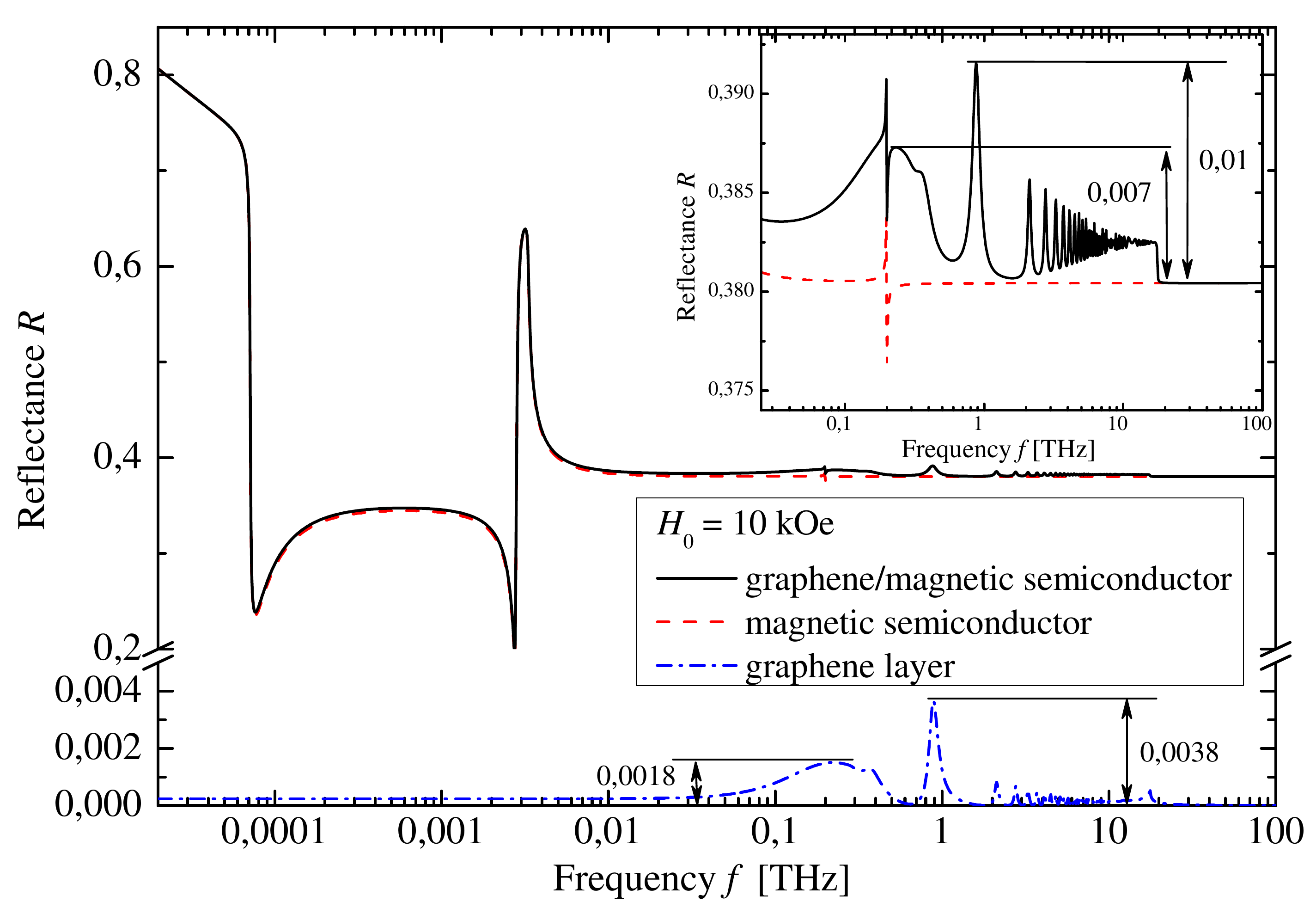}}
\caption{Comparison of frequency dependencies of electromagnetic waves reflectance $R$ for graphene - semi-infinite magnetic semiconductor structure, semi-infinite magnetic semiconductor and graphene layer. The value of external magnetic field is $H_{0}$ = 10 kOe. }
\label{ris:semi-inf}
\end{figure}

First of all, we will consider an influence of graphene layer on electrodynamic properties of  structure graphene - semi-infinite magnetic semiconductor. At the frequencies near the ferromagnetic resonant frequency spin oscillations will excited in magnetic semiconductor. Electromagnetic wave of such frequency will be strongly damped in the medium. This leads to  increase in reflectance. The energy of electromagnetic wave near the cyclotron resonance frequency will actively converse to the energy of the electrons rotating around the magnetic field lines. Thus, electromagnetic radiation of these frequencies will also strongly absorbed by medium. The resonant frequencies depend on external magnetic field value. Effect of graphene layers on the electromagnetic properties of the structure is more evident at the frequencies corresponding to carriers' transitions between Landau levels. At these frequencies, conductivity of graphene sharply increases, that leads to increase in reflectance. The results of numerical modeling are shown on Fig. \ref{ris:semi-inf}. One can see that combination of graphene layer with semi-infinite magnetic semiconductor leads not only to a simple algebraic sums of reflectances of the separate materials. Such peculiarity is caused by interaction of excitations of graphene layer with excitations of magnetic semiconductor. As seen from Fig. \ref{ris:semi-inf} such interaction leads to the fact that increase in reflectance caused by graphene carriers' transitions between Landau levels is about three times more for a structure graphene - magnetic semiconductor than for graphene layer only. Presence of graphene layer on the surface of magnetic semiconductor has almost no effect on the features of frequency dependence of reflectance, caused by ferromagnetic and cyclotron resonances. Note also, that calculated reflectance of graphene layer is very small (even at resonants it is less then 0.5 \%) that is in agree with other works \cite{17, 18, 49}.

\begin{figure} [h!]
\center{\includegraphics[width=90mm]{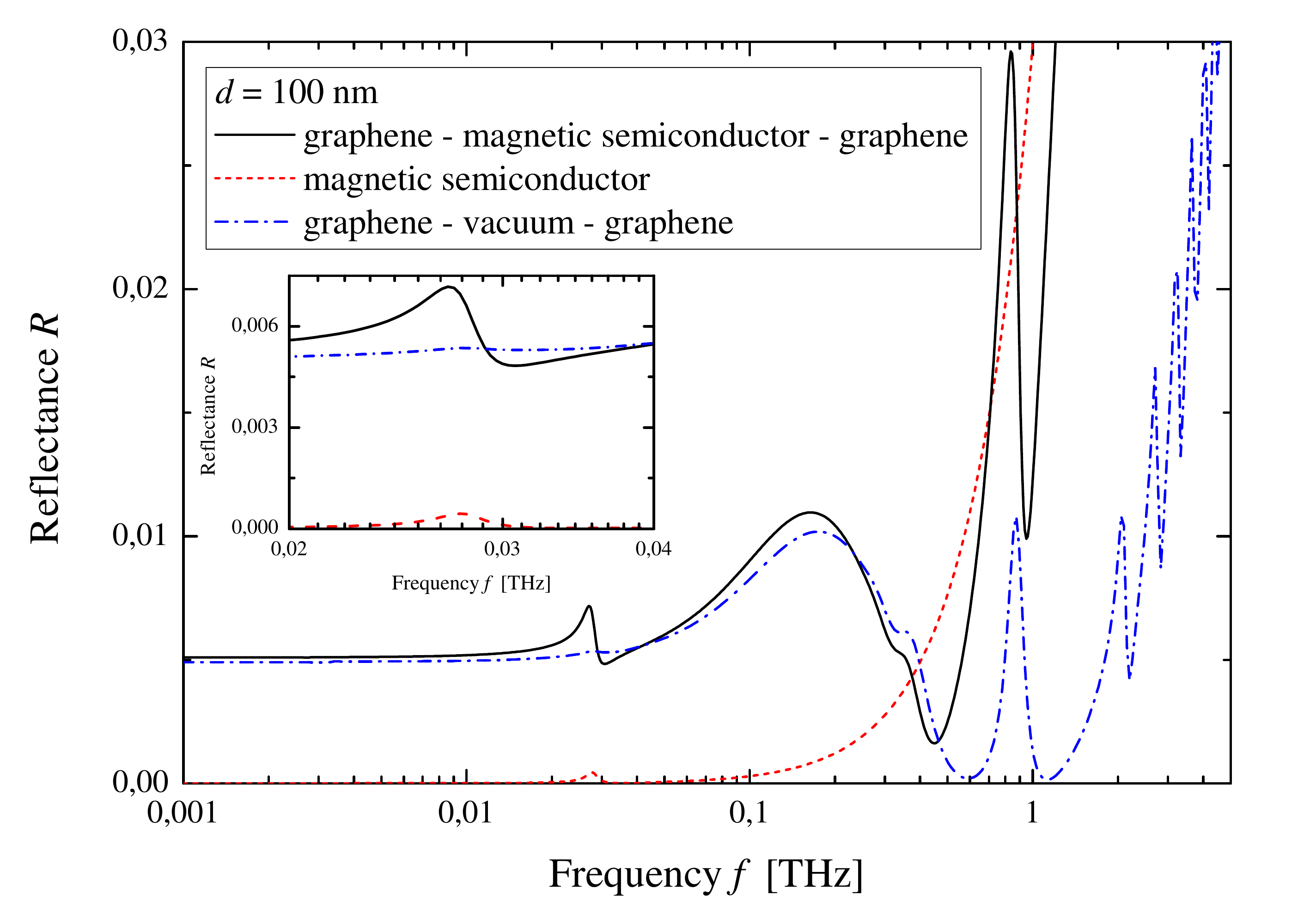}}
\caption{Comparison of frequency dependencies of electromagnetic waves reflectance $R$ for graphene -  magnetic semiconductor - graphene structure, magnetic semiconductor layer and graphene bilayer with thickness 100 nm. The value of external magnetic field is $H_{0}$ = 10 kOe. }
\label{ris:compare}
\end{figure}

Let us now consider influence of graphene layers on reflectance of sandwich-structure graphene - magnetic semiconductor - graphene. All features mentioned above will have a place also in this case. The main feature is the presence of dimensional resonances due to the interference of the waves reflected by the borders of the structure (graphene layers).  The results of numerical modeling are shown on Fig. \ref{ris:compare}. One can see, that reflectance of bilayer graphene is also very small (about 1 \% at resonances). Interaction between excitations in graphene and magnetic semiconductor leads to increase the resonances caused by graphene carriers' transitions between Landau levels much less then in case of semi-infinite magnetic semiconductor, but this interaction leads also to a sufficient increase in frequency of the resonance corresponding to cyclotron resonance in magnetic semiconductor. In common,  influence of graphene layers on reflectance of sandwich-structure graphene - magnetic semiconductor - graphene is less then 1 \%, but one has to expect more evident effect in multilayered structure based on graphene and magnetic semiconductor. In further, we will investigate only a features of electromagnetic waves reflection, transmission and absorption by the structure graphene - magnetic semiconductor - graphene, but the same features will have a place in case of multilayered structure.

\section{Reflection, transmission and absorption of electromagnetic waves by the structure}
\label{results}

\begin{figure} [h!]
\center{\includegraphics[width=90mm]{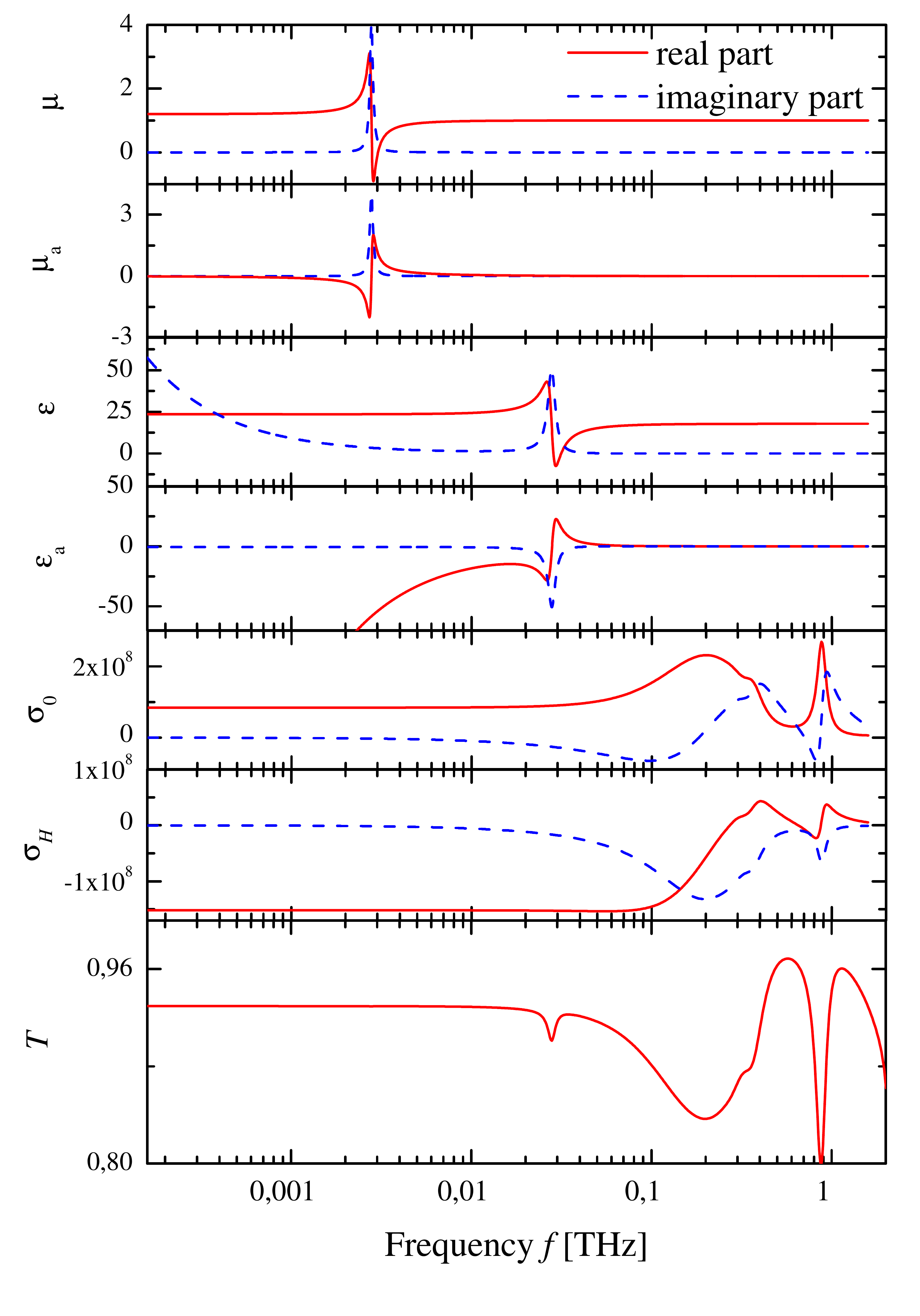}}
\caption{Frequency dependencies of permeability $\hat{\mu }$, permittivity $\hat{\varepsilon }$, and conductivity $\hat{\sigma }$ tensors components, and transmittance $T$. The thickness of magnetic semiconductor layer is $d$ = 100 nm, the value of external magnetic field is $H_{0}$ = 10 kOe.}
\label{ris:all}
\end{figure}

Due to resonant dependencies of components of permittivity, permeability tensors of magnetic semiconductor \eqref{eq:mu_tens}, \eqref{eq:eps_tens} and conductivity tensor of graphene \eqref{eq:sigma_tens}, it is clear that reflectance, transmittance and absorptance will have some features near the resonances. At the frequencies near the ferromagnetic resonant frequency $\omega \cong {{\omega }_{H}}{{\left( 1+{{\alpha }^{2}} \right)}^{-{1}/{2}\;}}$ spin oscillations (or magnons) will excited in magnetic semiconductor, and, therefore, electromagnetic wave will be strongly damped in the medium. This leads to decrease in transmittance and increase in reflectance and absorptance. Near the cyclotron resonance frequency $\omega \cong {{\omega }_{c}}{{\left( 1+{{\nu }^{2}} \right)}^{-{1}/{2}\;}}$ the energy of electromagnetic wave actively converse to energy of the electrons rotating around the magnetic field lines. Thus, electromagnetic radiation of these frequencies will also strongly absorbed by medium. Note that these effects will not only lead to features of reflectance, transmittance and absorptance, but also to change in the polarization of the reflected and transmitted waves. This is due to the fact that the wave of the polarization, which coincides with the direction of spin or electrons rotation, respectively, will be absorbed stronger. The resonant frequencies are determined by the value of external magnetic field. Effect of graphene layers on the electromagnetic properties of the structure is more evident at the frequencies corresponding to carriers' transitions between Landau levels. At these frequencies, conductivity of graphene sharply increases, that lead to increase in absorptance and reflectance (the structure becomes more "metallic") and to decrease in transmittance. These features can be seen in Figure \ref{ris:all}. 

\begin{figure} [h!]
\center{\includegraphics[width=90mm]{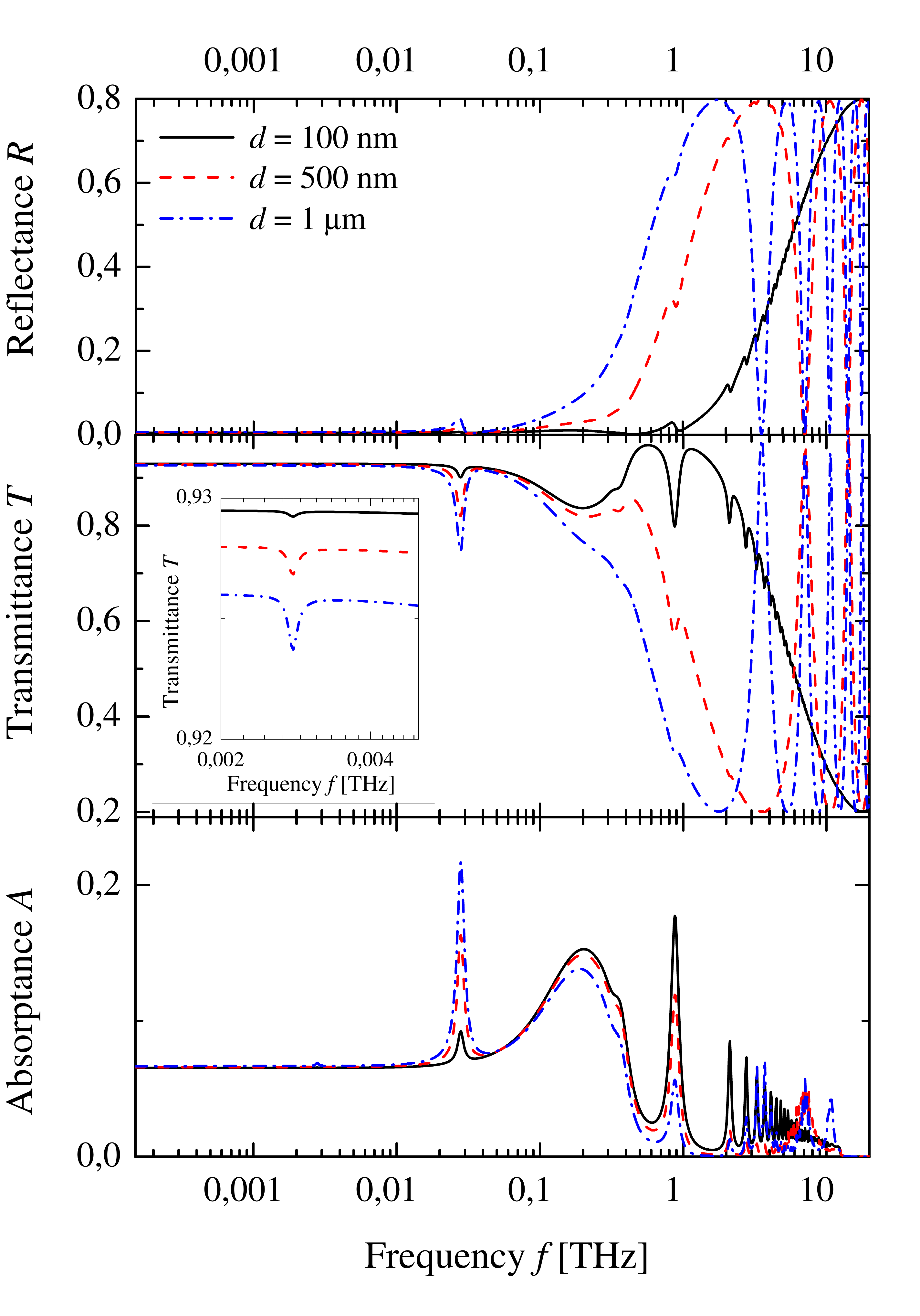}}
\caption{Frequency dependencies of electromagnetic waves reflectance $R$, transmittance $T$, and absorptance $A$ by graphene - magnetic semiconductor - graphene structure for different thicknesses $d$ of magnetic semiconductor layer. The value of external magnetic field is $H_0$ = 10 kOe.}
\label{ris:RTA_freq_thick}
\end{figure}

Calculation shows, that such effects are most evident in frequency dependencies of transmittance and absorptance. In the frequency dependence of transmittance shown in Figure \ref{ris:all} there is practically no feature associated with a ferromagnetic resonance. This is due to insufficient thickness of the magnetic semiconductor layer to observe a significant effect and due to selected material parameters. Increase in thickness of the structure leads to the fact, that the electromagnetic wave runs longer path in magnetic semiconductor and, consequently, more energy is transferred to magnetic subsystem. Figure \ref{ris:RTA_freq_thick} shows the results of frequency dependence of transmittance, reflectance and absorptance calculations for different thicknesses of the magnetic semiconductor layer. Additional to the effects mentioned above, one can also see that at frequencies $\omega <{{10}^{12}}$ s$^{-1}$ absorptance weakly depends on the thickness of magnetic semiconductor; hence, the absorption of the structure far from the resonances is mainly caused by graphene layers. At high frequencies, the contribution of size effects becomes significant. Characteristic size resonance frequencies for parameters \eqref{eq:parameters} are the following: $\omega \approx 2\cdot {{10}^{15}}$ s$^{-1}$ for thickness $d$ = 100 nm, $\omega \approx 4\cdot {{10}^{14}}$ s$^{-1}$ for thickness $d$ = 500 nm, and $\omega \approx 2\cdot {{10}^{14}}$ s$^{-1}$ for thickness $d$ = 1 micron.

It is well known \cite{29}, that values of the permeability tensor components at the resonant frequencies have a form
\begin{equation}\label{eq:resonant}
\begin{matrix}
  & \operatorname{Re}\left( {{\mu }_{res}} \right)=1+{{{\omega }_{M}}}/{2{{\omega }_{H}}}\;,\ \operatorname{Im}\left( {{\mu }_{res}} \right)={{{\omega }_{M}}}/{2\alpha {{\omega }_{H}}}\;, \\ 
 & \operatorname{Re}\left( {{\mu }_{a\ }}_{res} \right)=1,\ \operatorname{Im}\left( {{\mu }_{a\ }}_{res} \right)={{{\omega }_{M}}}/{2\alpha {{\omega }_{H}}}\;. \\ 
\end{matrix}
\end{equation}

So, the imaginary parts' peaks of the permeability tensor components, which are responsible for the absorption of electromagnetic waves by the material, are higher and narrower when the damping parameter of spin oscillations $\alpha$ is less. Choosing a medium with less damping parameter, one can observe a significant reflectance, transmittance and absorptance feature associated with the excitation of spin waves, even for thin films of magnetic semiconductors.

\begin{figure} [h!]
\center{\includegraphics[width=90mm]{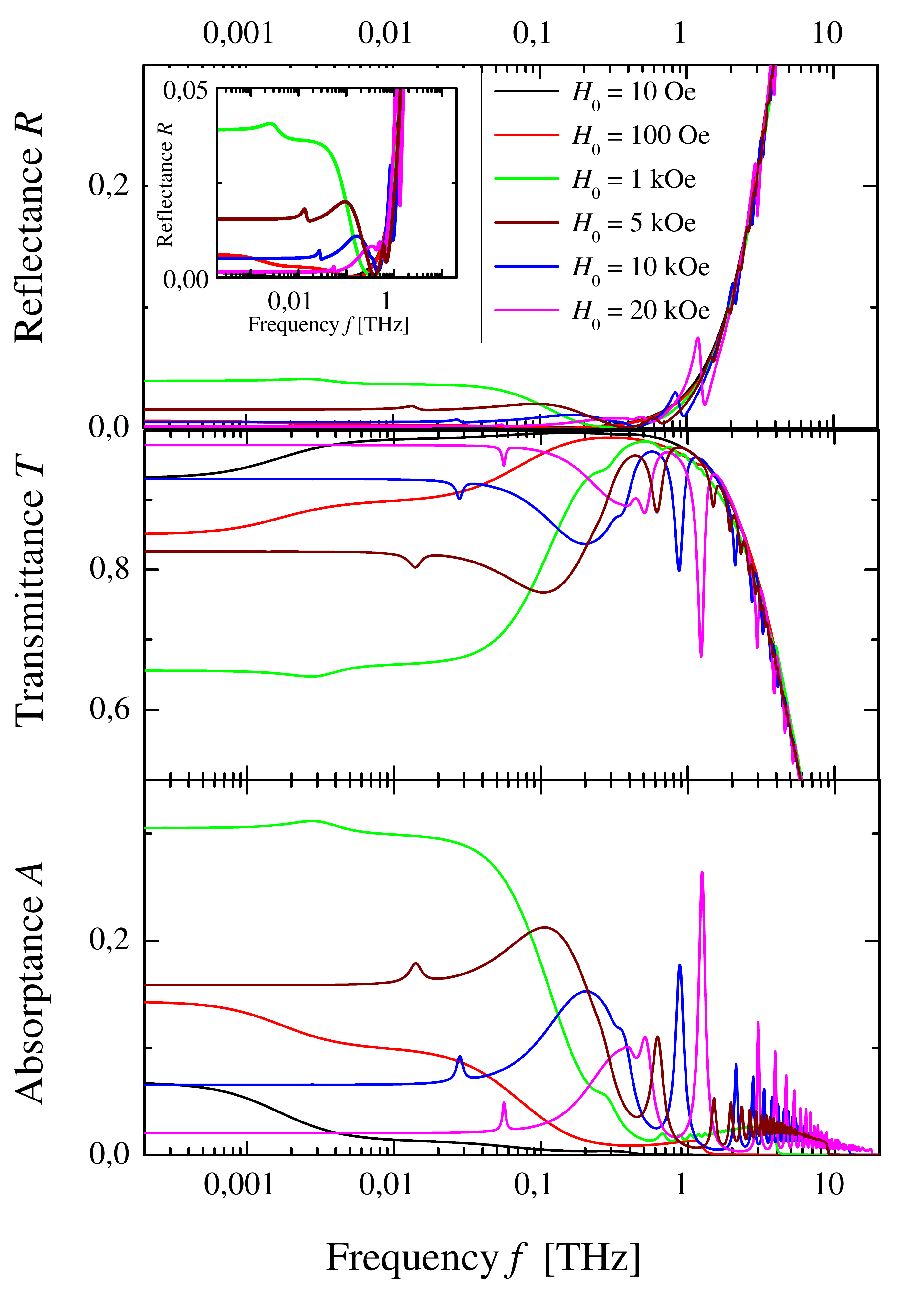}}
\caption{Frequency dependencies of electromagnetic waves reflectance $R$, transmittance $T$, and absorptance $A$ by graphene - magnetic semiconductor - graphene structure for different values of external magnetic field $H_0$. The thickness of magnetic semiconductor layer is $d$ = 100 nm.}
\label{ris:RTA_freq_field}
\end{figure}

Resonant frequencies values, energy of Landau levels in graphene, and distance between them depend on the value of external magnetic field. So, external magnetic field is an effective tool for controlling of electrodynamic characteristics of the structure. Figure \ref{ris:RTA_freq_field} shows frequency dependencies of reflectance, transmittance and absorptance of electromagnetic wave by graphene - magnetic semiconductor - graphene structure for different values of external magnetic field. One can see that at magnetic fields ${{H}_{0}}>50$ Oe at frequencies $\omega <{{10}^{13}}$ s$^{-1}$ the structure is completely transparent: reflectance is close to zero, absorptance is also very small. The feature at low frequencies mainly is caused by electrodynamic properties of magnetic semiconductor. For example, when ${{H}_{0}}$ = 10 Oe ${{\omega }_{H}}\sim {{10}^{8}}$ s$^{-1}$, ${{\omega }_{c}}\sim {{10}^{9}}$ s$^{-1}$ and the resonant frequencies are being below the frequency region under consideration. However, from \eqref{eq:resonant} it is clear, that at lower magnetic field magnitude of the imaginary part of permeability is more. The same is true for the components of permittivity tensor. This corresponds to greater absorption of electromagnetic waves energy by magnetic semiconductor. Increase in magnetic field value leads to shift of the resonances to higher frequencies, and this feature becomes more observable. Further increase in magnetic field value leads to increase in conductivity of graphene at frequencies $\omega <{{10}^{12}}$ s$^{-1}$, and, consequently, to increase in reflectance and absorptance and decrease in transmittance. At strong magnetic fields ${{H}_{0}}>10$ kOe graphene conductivity decreases and the resonant frequency of the magnetic semiconductor is shifted to higher frequencies, this leads to complete transparency of the structure at frequencies $\omega <{{10}^{12}}$ s$^{-1}$. At high frequencies transitions between Landau levels in graphene become essential, that leads to the jumps of conductivity. The change in the magnetic field leads to shift of the frequency and intensity of these jumps. 

\begin{figure} [h!]
\center{\includegraphics[width=90mm]{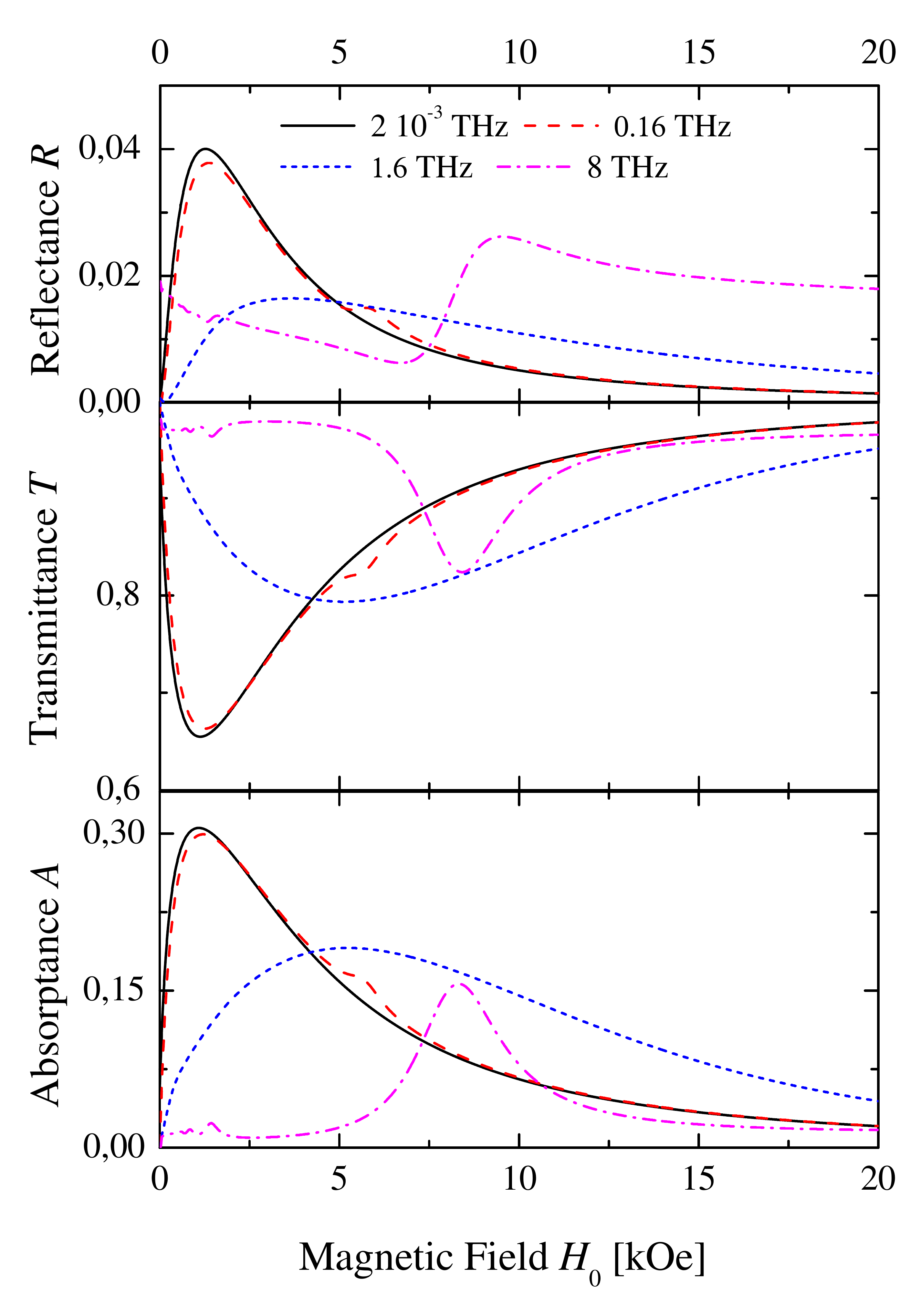}}
\caption{Field dependencies of electromagnetic waves reflectance $R$, transmittance $T$, and absorptance $A$ by graphene - magnetic semiconductor - graphene structure for different frequencies. The thickness of magnetic semiconductor layer is $d$ = 100 nm.}
\label{ris:RTA_field}
\end{figure}

Figure \ref{ris:RTA_field} shows field dependencies of reflectance, transmittance and absorptance of electromagnetic wave by graphene - magnetic semiconductor - graphene structure. The field dependencies have a resonant form. Increase in frequency leads to higher resonant values of magnetic field. A large peak at sufficiently low frequencies $\omega <{{10}^{12}}$ s$^{-1}$ is associated with the change of graphene conductivity due to applying magnetic field. In the medium-frequency region $\omega \sim {{10}^{12}}$ s$^{-1}$ a small secondary peak associated with the shift of the cyclotron frequency is observed. At certain magnetic field value this peak enters in considered frequency domain. At lower frequencies, this peak is not observable, because electromagnetic properties of the structure at these frequencies are mainly determined by graphene layers; cyclotron resonance peak is lost against graphene contribution. At higher frequencies, this effect is absent; this is due to the narrow width of the considered range of magnetic fields. At frequencies higher than $\omega \sim {{10}^{13}}$ s$^{-1}$ , several smaller resonances become observable besides large resonance. These resonances are caused by the irregular graphene conductivity associated with transitions between Landau levels. 

\section{Conclusions}
\label{conclusion}
Investigation of dynamic characteristics of graphene - magnetic semiconductor - graphene sandwich-structure placed in an external magnetic field showed that reflectance, transmittance and absorptance of electromagnetic waves of such structure can be efficiently controlled. For structure of sufficiently small size, that have a practical interest, the structure is weakly reflective. Transmittance of electromagnetic waves can be changed with the change of external magnetic field. At microwave range of frequencies, this change may reach 35-40 \% (from 100 \%) in weak magnetic fields 1-2 kOe. At terahertz frequencies the change is not so great, but also significant: up to 20-25 \% at magnetic field values of 8-10 kOe. Not passing through the material energy of electromagnetic radiation is mainly absorbed by the structure. Absorptance of the structure can be controlled from 0 \% to 30 \% at frequencies $10^{10} < \omega < 10^{12}$ s$^{-1}$ and from 0 \% to 15 \% at frequencies about $5 \cdot 10^{13}$ s$^{-1}$ by external magnetic field values of $0 < H_0 <10$ kOe at thickness value about 100 nm. Reflectance, transmittance and absorptance also depend on the thickness of the structure, which mainly affects values of $R$, $T$, $A$, at frequencies near the ferromagnetic and cyclotron resonances.

Note that permeability tensor of magnetic semiconductor \eqref{eq:mu_tens}, \eqref{eq:eps_tens}, and conductivity tensor of graphene \eqref{eq:sigma_tens} are antisymmetric. This leads to change in the polarization of the electromagnetic wave. In the general case, both reflected and transmitted waves are elliptically polarized. The polarization of reflected and transmitted waves can also be controlled by an external magnetic field value.

Let us note now some features that can make a difference, but were not considered in this paper. For example, it is influence of the demagnetizing field. It is well known \cite{29}, that in case of an infinitely thin plate, taking into account demagnetizing field leads to decrease in the ferromagnetic resonance frequency on value about ${{\omega }_{M}}$. So, to observe the effects associated with this resonance will be required several large magnetic fields. The study of oblique incidence of the electromagnetic wave on the structure is also interesting. In this case, waveguide modes can be excited between graphene layers. Interest for further research is also multilayered graphene - magnetic semiconductor structure. Increasing the number of graphene layers should lead to greater effects associated with resonances of graphene conductivity, to increase in the frequency range of these effects, and to shift of resonances. Some of these features have been recently demonstrated in \cite{50}.

\section{Acknowledgements}
\label{acknowledgements}
The reported study was supported in part by RFBR, research project No. 13-07-00462.






\end{document}